\documentclass[useAMS,usenatbib]{mn2e}
\usepackage[dvips]{graphicx}

\title{X-Ray Spectral Variations of U Gem from Quiescence to Outburst}
\author[T. G\"{u}ver, C. Uluyaz\i, M. T. \"{O}zkan and E. G\"{o}\u{g}\"{u}\c{s}]{T. G\"{u}ver
$^{1}$\thanks{E-mail:tolga@istanbul.edu.tr}, C. Uluyaz\i$^{1}$, M. T.\"{O}zkan$^{2}$ and E. G\"{o}\u{g}\"{u}\c{s}$^{3}$
\\
$^{1}$Istanbul University, Science Faculty Department of Astronomy \& Space Sciences, 34119, Turkey \\
$^{2}$Istanbul University, University Observatory, 34119 Turkey\\
$^{3}$Sabanc\i~University, FENS, 34956 Turkey}
\begin{document}

\date{}

\pagerange{\pageref{firstpage}--\pageref{lastpage}} \pubyear{2006}

\maketitle

\label{firstpage}

\begin{abstract}
In this paper we report the discovery of a high energy component of the X-ray spectra of U Gem , 
which can be observed while the source is in outburst. We used Chandra and XMM-Newton observations to compare 
the quiescence and outburst X-ray spectra of the source. The additional component may be the result of the reflection of 
X-rays emitted from an optically thin plasma close to the white dwarf, from the optically thick boundary layer 
during the outburst. Another possible explanation is that some magnetically channeled accretion may occur onto the equatorial 
belt of the primary causing shocks similar to the ones in the intermediate polars as it was suggested by \citep{w2002}. 
We have also found a timing structure at about 73 mHz ($\sim$13.7 s.) in the RXTE observation, resembling dwarf novae oscillations (DNOs).
\end{abstract}


\section{Introduction}

Dwarf novae are highly variable mass-exchanging binary star systems containing a white dwarf and
a late-type K or M star. They are characterized by occasional outburst episodes
which typically take place on a timescale of 100 days and last for about 15 days. 
Outbursts are believed to be triggered by a thermal instability in the accretion disk that drives 
the disk from a low-temperature, low mass accretion rate, quiescent state to a hot, high-$\dot{\rm{m}}$ 
outburst state. 

About every 120 days prototypical dwarf nova U Gem exhibits outbursts, where the V magnitude of 
the system rises from $\sim14$ to $\sim9$ \citep{s1984}. 
During outbursts U Gem is a bright EUV source. This emission is interpreted as optically thick radiation 
from a $\sim140.000 \rm{K}$ boundary layer \citep{l1996}. Besides, there is observational evidence 
that the observed energy from the boundary layer and that from the disk is comparable \citep{l1996} 
which is in accordance with the simple mass accretion scenario \citep{p1977}. 
In quiescence the UV spectra of U Gem is dominated by the white dwarf emission as observed
with IUE \citep{k1991}, Hopkins Ultraviolet Telescope (HUT) \citep{l1993} and 
Hubble Space Telescope (HST) \citep{l1994}. 
Studies of the white dwarf in U Gem indicate evidence for an equatorial accretion belt \shortcite{c1997},
\shortcite{l1993} which is cooling during the quiescence interval. 

X-ray emission (0.2 - 10 keV) has been observed from U Gem both in quiescence and outburst \shortcite{s1978}, 
\shortcite{c1984}, \shortcite{s1996}.  
Unlike other well known dwarf novae systems, like SS Cygni, hard X-ray flux of U Gem does not decrease in outbursts \citep{m2000}. 
The fact is that the increase in the hard X-ray (2-15 keV) flux is less than that of the soft X-ray flux (0.1-4 keV) \shortcite{c1984}, \shortcite{m2000}. 
Chandra observations of U Gem in quiescence indicate that the hard X-ray emission arises from a gas with a small scale height ($<10^{7}cm$) close to the white dwarf \citep{s2002}, (hereafter S02).

Here we present detailed hard X-ray emission properties of U Gem in outburst phase for the first time. 
We find an extra non-thermal X-ray component arising during outburst which can be intrepreted as a sign of a transient magnetosphere occuring in the boundary layer of U Gem as a result of the increase of the mass accretion rate \citep{w2002}.

In section 2, we present observations and our data analysis. In sections 3 and 4, we show observational
differences between quiescence and outburst phases, respectively. In section 5, we discuss these results.

\section{Observations and Data Analysis}

\begin{figure*}
\centering
   \includegraphics[width=8cm,height=4cm]{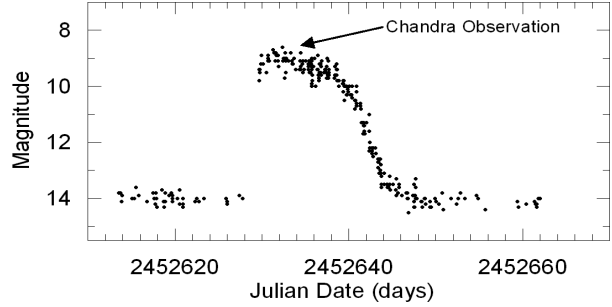}
   \caption{AAVSO lightcurve of the optical outburst in 2002. The arrow marks the time of Chandra observation.}   
   \label{obs}
\end{figure*}

We use four pointed observations of U Gem with three satellites namely Chandra,
XMM-Newton \citep{j2001} and RXTE. These observations enable us to study 
both the outburst and the quiescence states of the source. 
Details of these observations can be found in Table \ref{obs_list}. We should note that although both outbursts are normal outbursts, the outburst in 2004 is a few days longer than the outburst in 2002. Second Chandra observation was made during the peak of a normal outburst in 2002 (see Figure \ref{obs}), while in 2004, a series of RXTE observations were made in order to cover the whole outburst.
Time of the Chandra X-ray observation is marked on the AAVSO data of the outbursts 
in Figure 1.

\begin{table*}
\caption{Details of all U Gem observations used in this study.}
\centering
\begin{tabular}{lccccc}
\hline
Satellite  & Obs ID     & Appr. Exp.  & Ins. / Grating& Start Time            & Source State \\
           &            & (ks)        &               & (UT)                  & \\
\hline
Chandra    & 647        & 100       & ACIS-S / HETG   & 2000-11-29 12:00:17   & Quiescence \\
XMM        & 0110070401 & 23        & All             & 2002-05-09 10:46:06   & Quiescence \\ 
Chandra    & 3767       & 67        & ACIS-S / HETG   & 2002-12-26 09:27:36   & Outburst \\
RXTE       & 80011-01   & 115       & PCA             & 2004-02-27 12:49:54   & Outburst \\
\hline
\end{tabular}
\label{obs_list}
\end{table*}

In the public data archive of Chandra there are three observations of U Gem with a total approximate exposure time 
of 217.000 seconds. For this study, however, we won't present the LETG observation since we will be interested 
mainly on the high energy X-ray emission of the source. Chandra data was analysed by Chandra Interactive 
Analysis of Observations\footnote{CIAO, http://cxc.harvard.edu/ciao/} software version 3.2
and Chandra Calibration Database (CALDB) version 3.0.0. A new bad pixel file 
was created to identify and flag hot pixel and afterglow events in ACIS observations. 
This tool searches for, pixels where the bias value is too low or too high, classifies the events on 
suspicious pixels as being associated with cosmic ray afterglows, hot pixels or astrophysical 
sources and add newly found bad pixels to the output new bad-pixel file. 
The first Chandra observation, while the source was in a quiescence state, was reported in detail by S02.

For the extraction of the scientific information from the XMM-Newton data, XMM-SAS version 6.1 and the latest 
available calibration files were used. We used {\it epproc, emproc} meta tasks to extract calibrated 
source events. Since we will be interested in mainly the high energy part of the spectrum, 
we will not be presenting RGS data, which was presented by \citep{p2005}. 

RXTE pointed observations of U Gem were performed between 27 February and 14 March 2004, with a total
effective exposure time of 115 ks. Onboard RXTE, there are two main instruments; the Proportional
Counter Array (PCA), an array of five nearly identical Proportional Counter Units (PCU) that are
sensitive to photon energies between 2 - 60 keV, and the High Energy X-ray Timing Experiment (HEXTE)
that is sensitive 20 - 200 keV photons. In this study we only used data collected with the PCA.
For each RXTE observation, we extracted spectrum using the Standard2 data collected with PCU2
since it was operational in all the 24 RXTE pointings. The background spectrum for each pointing
was obtained using the PCA background models for bright sources.

We have used XSPEC v11.2 \citep{a1996} for the analysis of continuum spectra. In order to use Chi Square Analysis we have 
grouped all spectra to have at least 40 counts in each bin. Since we have high enough count rates in grating spectra, 
we did not try to fit the zero-order ACIS spectra therefore we did not have to model the pile-up effects of the ACIS CCDs which was also mentioned in S02 for the quiescence data.
Because of the difference in the responses of detectors we took different energy ranges for different datasets, 
for Chandra gratings we used 0.5 - 7.0 and 0.8 - 7.5 
keV energy range for MEG and HEG, respectively, for the XMM-Newton EPIC-PN \citep{t2001} data we used 
the 0.2 - 10 keV region.

\section{Quiescence}
\begin{figure*}
\centering
\rotatebox{270}{
   \includegraphics[scale=0.5]{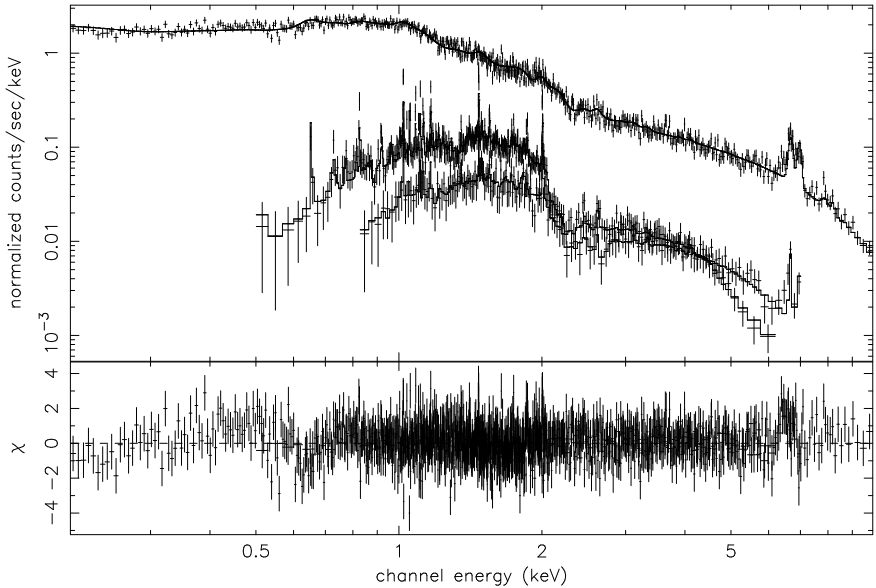}}
   \caption{Best fit model to all observations of U Gem during quiescence.}  
   \label{quiescence}
\end{figure*}

During quiescence we have two observations of U Gem from two different satellites. 
We simultaneously model X-ray spectra from these pointings assuming that
the system was in the same X-ray regime during these observations, as it was the case
for the optical emission ($\sim14\rm{mag}$).
We fit the spectrum with the so called XSPEC models cemekl or cvmekl. 
These models assume an optically thin plasma cooling from a maximum temperature that follows a 
power-law distribution. The only difference between the cemekl and cvmekl models is that, 
the latter one calculates the abundances of all element with respect to the solar values, and the other one 
calculates these values for 13 most abundant elements separately. An application of this model and a more detailed 
discussion can be found in \citep{p2003}. With a little worse $\chi_{\nu}^{2}$, value we could also fit the spectrum with another 
XSPEC model mkcflow which was also used for CVs by \shortcite{m2003}. 
Our results for the quiescence state are presented in Table \ref{qspec} and the spectra can be seen in Figure \ref{quiescence}. 
For the calculation of N$_{\rm{H}}$ and abundances we used XMM EPIC-PN and Chandra MEG data as a reference to the other 
data sets. From the narrow lines in the Chandra spectra of U Gem in quiescence, a low-velocity emission region close 
to the white dwarf was suggested by S02. 

\begin{table*}
\centering
\caption{Best fit model parameters for quiescence. Uncertainties are calculated for 90 \% confidence
interval. Fluxes are calculated for the given energy ranges of the datasets. Abundance values are fixed to the
value obtained from MEG fit. NH values are fixed the the value obtained from EPIC-PN fit.}
\begin{tabular}{lcccccccc}
\hline
Dataset  & Model  & NH              & $\rm{T_{max}}$ & Power-Law Index & $\dot{\rm{m}}$&Flux & Abundance & $\chi_{\nu}^{2}$ \\
         & & $10^{20} \rm{atoms/cm^{2}}$ & kT (keV)  & $\alpha$  & $10^{-12}$ gr/yr  & $10^{-11}\rm{ergs/cm^{2}/s}$ & $\odot$   & \\
\hline
HEG      & cemekl & - & 22.47$\pm7.15$ & 0.96$\pm0.27$  &- & 1.15$\pm0.05$  & - & 1.006 \\
MEG      & cemekl & - & 23.22$\pm7.70$ & 1.05$\pm0.14$  &- & 1.05$\pm0.03$  & 1.30$\pm0.22$ & 1.006\\
EPIC-PN  & cemekl & 0.98$\pm0.26$      & 29.25$\pm4.36$& 0.87$\pm0.06$ &- & 0.78$\pm0.02$  & - & 1.006 \\
HEG      & mkcflow & - & 28.41$\pm3.80$& -& 6.64$\pm0.36$ &1.14$\pm0.1$ & - & 1.093\\
MEG      & mkcflow & - & 35.08$\pm2.78$&- & 5.42$\pm0.15$ &1.16$\pm0.04$ & 1.31$\pm0.10$& 1.093\\
EPIC-PN  & mkcflow & 0.47$\pm0.19$ & 27.49$\pm0.96$& -& 4.77$\pm0.06$ &1.17$\pm0.02$ &-& 1.093\\
\hline
\end{tabular}
\label{qspec}
\end{table*}

\section{Outburst}

One of the interesting properties of the X-ray spectra during outburst is the broadened emission lines, 
for the quiescence phase properties of the X-ray emission lines have been studied in S02 
and from the measured broadening of the lines it has been suggested that the lines originate
from a low velocity material instead of an inner disk region rotating at the Keplerian velocity. From the
investigations of the quiescent spectra, narrow X-ray emission lines appear to be a common 
situation for some other CVs also. However outburst X-ray spectrum of U Gem shows that most of the 
emission line fluxes are increased and most of them are broadened. To be able to make a comparison, 
in Table \ref{emis_line} we have given a summary of some of the emission lines detected in both 
observations of Chandra. 

In order to model the emission lines, listed in Table \ref{emis_line}, we first fit the local region of a line with a polynomial function so that the continuum can be estimated. We than add a gaussian to fit the residuals from the continuum. ATOMDB database is used for the identification of the lines, and the calculation of the velocities are done by assuming, the measured full width at half maximum (FWHM) values of the emission lines arise only because of the Doppler broadening of the material in a Keplerian orbital motion around the white dwarf. 

\begin{table*}
\centering
\caption{Properties of the emission lines, which were observed both during quiescence and outburst, are
given. All the line detections were made by Medium Energy Grating, MEG, onboard Chandra. Velocities of the
lines are not corrected to the orbital inclination of the system. Errors for some of the lines velocities are not given because these lines are modelled by fixing the FWHM value to 0.023 \AA~which is the limit for MEG.}
\begin{tabular}{lccccc}
\hline
Line  & Wavelength & Quiescence Flux   & Outburst Flux     & Quiescence Velocity & Outburst Velocity \\
      &  \AA    & $10^{-5} \rm{Photons/cm^{2}/s}$& $ 10^{-5} \rm{Photons/cm^{2}/s}$ & km/s   & km/s  \\
\hline
S XV  & 5.0665     & 0.94$\pm0.52$  & 4.16$\pm1.47$  & 1350                & 4180$\pm2000$ \\
Si XIV& 6.1804     & 2.17$\pm0.33$  & 6.91$\pm0.80$  & 1120                & 3080$\pm460$  \\
Mg XII& 8.4192     & 1.86$\pm0.27$  & 2.59$\pm0.49$  & 820                 & 2200$\pm690$  \\
Mg XI & 9.1687     & 0.70$\pm0.26$  & 1.26$\pm0.62$  & 552$\pm277$         & 2000$\pm900$  \\
Fe XXIV& 10.6190   & 2.14$\pm0.43$  & 4.89$\pm0.96$  & 1340$\pm440$        & 4450$\pm1000$ \\
Ne X   & 12.1321   & 3.46$\pm0.65$  & 11.2$\pm1.85$  & 1000$\pm290$        & 3450$\pm550$  \\
Fe XVII& 15.0140   & 3.17$\pm0.89$  & 12.3$\pm2.49$  & 740$\pm350$         & 2470$\pm760$  \\
O VIII & 16.0055   & 1.14$\pm0.70$  & 3.98$\pm1.87$  & 430                 & 1400$\pm1200$ \\
Fe XVII& 17.0510   & 2.97$\pm1.10$  & 17.9$\pm3.96$  & 400                 & 2830$\pm913$  \\
\hline
\end{tabular}
\label{emis_line}
\end{table*} 

\begin{figure*}
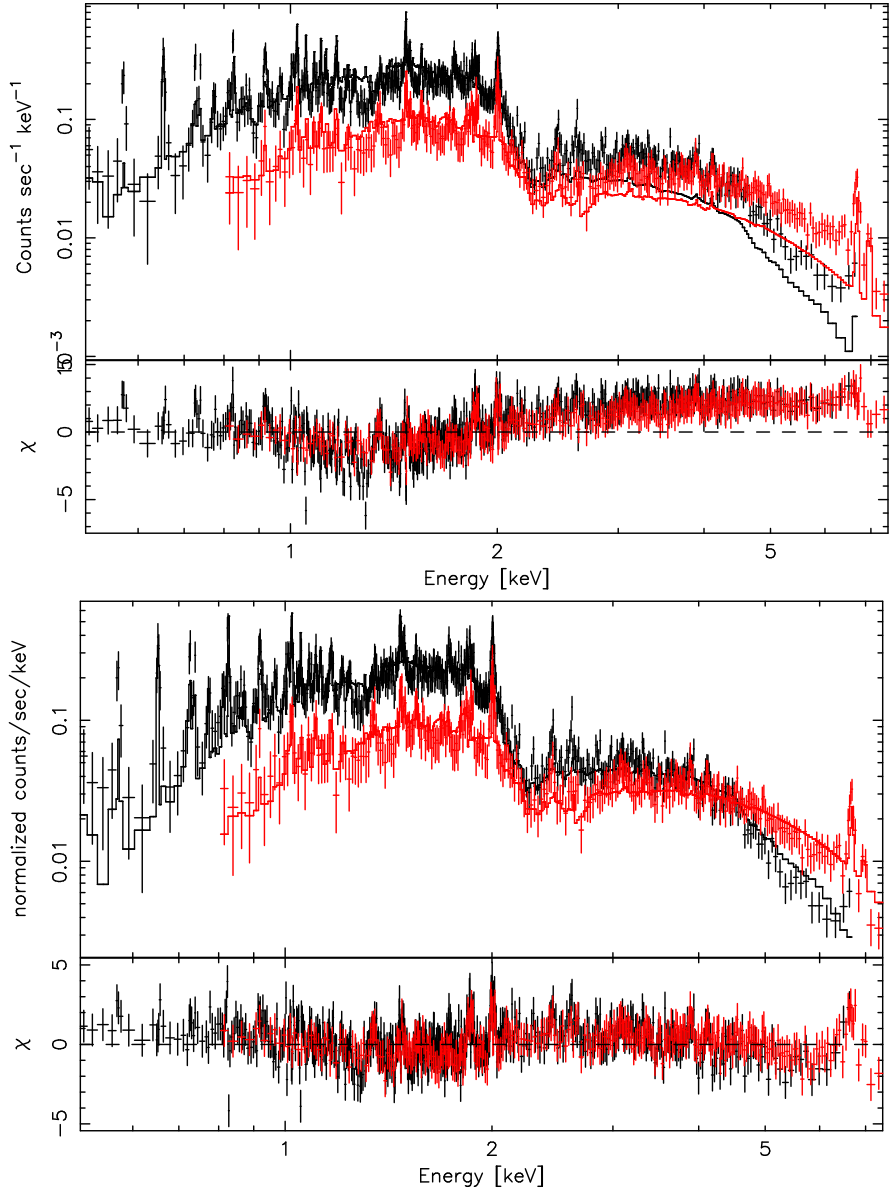

\rotatebox{270}{
   \includegraphics[scale=0.50]{figure_03_1.ps}
   \includegraphics[scale=0.50]{figure_03.ps}}
   \caption{Outburst spectra fitted with the cemekl model, upper panel shows the results when the best fit parameters for the quiescence used. Lower panel presents the results when a power-law model added to the cemekl model.}
\label{outburst_spectra}
\end{figure*}

Interestingly, we find excess high energy component (as seen in the upper panel of Figure \ref{outburst_spectra})
in the outburst spectra when fitted with the model adequately represent the quiescence
spectrum. This did not fit the data neither with the frozen model parameters (except flux, which was found as $\sim 3.08 10^{-11}\rm{ergs/cm^{2}/s}$) found from the quiescence phase analysis ($\chi_{\nu}^{2} \sim 2.95$) (see upper panel of Figure \ref{outburst_spectra}) nor with allowing the parameters to vary ($\chi_{\nu}^{2} \sim 1.877 $), which also gave physically unacceptable
values (see Table \ref{outburst_param}). 

In order to resolve this, we fixed the cemekl model parameters as obtained from quiescence 
and added a power-law component to this model. This way we could obtain an acceptable fit with a $\chi_{\nu}^{2}$ value of 1.41. 
Then we set the parameters to be free which reduced the $\chi_{\nu}^{2}$ even further to 1.3. Results of these fits are given 
in Table \ref{outburst_param}, and the spectra can be seen in bottom panel of Figure \ref{outburst_spectra}. We have also tried some other models, like thermal-bremmstrahlung, which either gave unphysical results or unacceptable $\chi_{\nu}^{2}$ 
values. Just for presentation, these results are given in Table \ref{outburst_param}.

We have also analysed all individual RXTE/PCA spectra obtained
during the 2004 outburst. We fitted the background subtructed PCA spectrum in
3$-$20 keV range with a thermal (cemekl) plus power law model but fixing the
parameters of the thermal component at the values obtained from Chandra
observations in 2002. As a result, we find that the hard X-ray spectral
shape of U Gem vary significantly throughout the 2004 outburst; in 14 of
23 spectra, a power law model with indices between 0.28 and 1.34 were needed
to succesfully fit the data, while in 9 of 23 spectra, no power law component
was required. This trend provides evidence for transient spectral changes
in U Gem during the outburst episode. However, it is crucial to note that
the PCA response does not allow to constrain the low energy (thermal)
component and the projected Chandra spectral shape in 2002 may not reflect
the intrinsic spectral shape in 2004. To conclusively quantify the high
energy spectral variations, simultaneous wide band X-ray observations
are needed.

\begin{table*}
\centering
\caption{Best fit model parameters for only outburst. Uncertainties are calculated for 90 \% confidence interval. Abundance is also fixed to 1.05 solar. Additional model is power-law whose photon index is given with the $\Gamma$. N$_{H}$ values are fixed to the value obtained from MEG fit.}
\begin{tabular}{lcccccccc}
\hline
Dataset  & Model        &NH                     & $T_{Max}$   & Power-Law index   & Photon Index & T    &Flux                   &$\chi_{\nu}^{2}$\\
         & Name         &$10^{21} \rm{atoms/cm^{2}}$ & kT (keV)    & of cemekl $\alpha$& $\Gamma$     & keV  &$10^{-11}\rm{ergs/cm^{2}/s}$&           \\
\hline
MEG      & cemekl       & 0.37$\pm0.004$       & 100$\pm3.4 $  &0.52$\pm0.04$      &  -        &  -    &3.00   & 1.877 \\
HEG      & cemekl       & 0.37$\pm0.004$       & 100$\pm13.2$  &0.69$\pm8.23$      &   -      &  -    &3.40    & 1.877 \\
HEG      & cemekl+pow   & 0.45$\pm0.45$        & 9.49$\pm4.1$  &1.35$\pm0.49$& 0.54$\pm0.16$&  - &3.31$\pm1.38$& 1.297 \\  
MEG      & cemekl+pow   & 0.45$\pm0.45$        & 18.44$\pm10.5$&0.26$\pm0.11$& 0.64$\pm0.06$&  -    &3.39$\pm0.53$& 1.297 \\
HEG      & cemekl+bremss& 3.43$\pm0.38$        & 94.17         & 0.07              &    -          & 199.3& - &- \\
MEG      & cemekl+bremss& 3.43$\pm0.38$        & 1.73          & 0.01              &     -         & 199.3& - &-\\
\hline
\end{tabular}
\label{outburst_param}
\end{table*}

\subsection{Quiescent Subtracted Spectrum}

In order to investigate the difference in the spectral properties of the quiescence and outburst phases 
more reliably we have employed the following technique: we assumed the spectrum of the quiescent phase
as that of the background for the outburst spectrum so that the remaining spectral information would be 
only due to spectral changes in the source during outburst.

For the fitting we couldn't use the lower energies because, when subtracted, count rates of these regions were too
low to make a reliable fit so we have used 1.0-6.5 keV and 1.0-7.0 keV range for MEG and HEG, respectively.

\begin{figure*}
\centering
\rotatebox{270}{
	\includegraphics[scale=0.5]{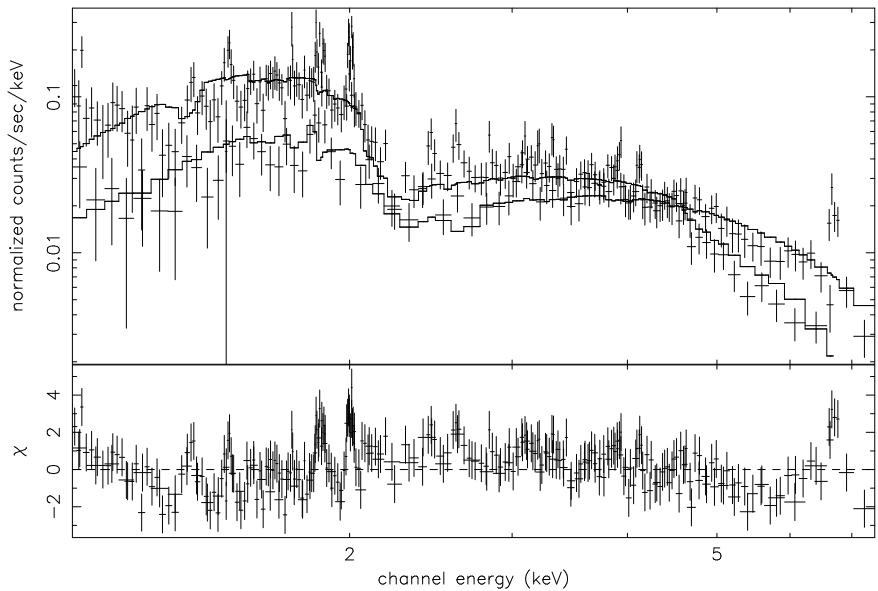}}
   \caption{Quiescent subtracted, outburst spectra and the best fit model cemekl+pow.}  
\label{subtracted}
\end{figure*}

Resultant spectrum can be best fit by a power-law with a $\chi_{\nu}^{2}$ value of 0.98.
By only fitting a power-law to the subtracted data, the change in the X-ray emission lines can also be clearly seen 
which have more flux and are appearently broader than in quiescent state.
In order to investigate the origin of the difference and to fit the emission lines, we have also tried some other 
XSPEC models like the derivatives of the mekal models and bremss model, these fittings either gave very bad $\chi_{\nu}^{2}$ values or unphysical results. Results are given in Table \ref{sub} and the spectrum can be seen in Figure \ref{subtracted}.

\begin{table*}
\centering
\caption{Results of the modelling of the quiescence subtracted spectra. Model is power-law. 
         Fluxes are calculated for the 1.0 - 7.0 keV range, because of low signal to noise ratio below 1 keV. 
	 N$_{H}$ is frozen to the value obtained from NH calculator of HEASARC.}
\begin{tabular}{lccccc}
\hline
Dataset & Model Name &  $T_{Max}$   & Photon Index & Flux  		 & $\chi_{\nu}^{2}$ \\
        &            &  keV         &$\Gamma$	 & $10^{-11}\rm{ergs/cm^{2}/s}$&	       \\
\hline
MEG     & pow        & -	    &0.46$\pm0.06$ & 2.42$\pm0.14$ 	 & 1.02   \\
HEG     & pow        & - 	    &0.67$\pm0.08$ & 2.25$\pm0.26$ 	 & 1.02   \\
MEG     & mekal      &79.89$\pm3.40$& -             & 1.50$\pm0.03$	 & 1.68   \\
HEG     & mekal      &79.89$\pm5.11$& -             & 1.98$\pm0.05$	 & 1.68   \\
MEG     & bremss     &199.4$\pm15.2$& -             & 1.52$\pm0.01$       & 1.62  \\
HEG     & bremss     &199.4$\pm23.8$& -             & 1.98$\pm0.01$       & 1.62  \\
\hline
\end{tabular}
\label{sub}
\end{table*}

\subsection{Timing Analysis}

To examine the timing characteristics of the hard X-ray emission from U Gem during
the February-March 2004 outburst, we performed a standard timing analysis as follows. 
We generated the 2 - 15 keV light curves with time binning of 16 ms using the PCA event 
mode data and converted the times to the Solar System Barycenter. For each RXTE pointing, 
we divided light curves into 512 s long segments. We then applied Fourier transformation 
to each of 512 s data segment and computed associated Fourier (Leahy) powers. We obtained 
the power spectrum of each pointing by averaging the Fourier power spectra of all available 
segments. 

\begin{figure*}
\centering
\rotatebox{0}{
	\includegraphics[scale=0.5]{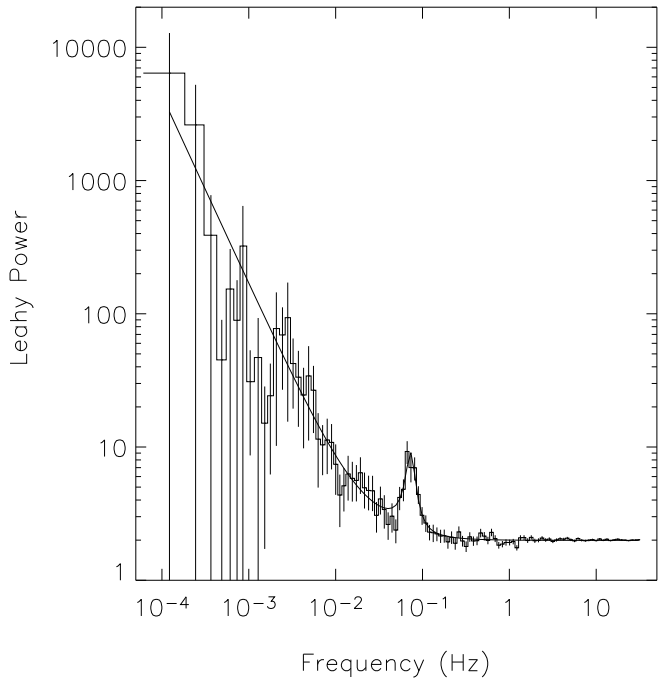}}
   \caption{Power spectrum of the PCA data obtained on 2453077 JD, a QPO structure can be seen near 73 mHz.}  
\label{timing}
\end{figure*}

We find that the power spectra were dominated by red noise structure below about 10 mHz and 
consistent with powers due to Poisson count fluctuations at frequencies above. We determine
that the RMS amplitudes of low frequency fluctuations vary in the range of 1.2\% and 7.3\%. 
Interestingly, we detect a quasi periodic timing structure in the RXTE observations on 2004 
March 13. The power spectral density of this pointing is shown in Figure \ref{timing}. We fitted the 
power spectrum with the sum of a constant, a power law and a Lorentzian function to account
for the Poisson noise, low frequency red noise and the QPO structure, respectively. For the
QPO feature, we obtain the peak frequency as 73$\pm$9 mHz and RMS amplitude as 1.0$\pm$0.3\%.
The peak frequency of this oscillation corresponds to 13.7 s.

\section{Discussion}

In this work we have investigated U Gem's X-ray behaviour during outburst. We found that unlike quiescence, one 
cannot fit the X-ray spectrum of U Gem with a simple mekal based multi-temperature spectral model. Instead, an
additional non-thermal component, most likely a power-law, is needed. We should also note that when fitting the excess hard X-ray emission with a power-law, deviations from the continuum can be seen (see Figures \ref{subtracted} and \ref{outburst_spectra}). We think that these deviations arise because of the broadening and increase of the fluxes of the X-ray emission lines when compared to the quiescence phase. In order to be able to obtain better fits to the outburst data one should also be able to model the broadening of the lines physically, which is currently unavailable.

Interpretation of the extra X-ray component can be made as follows: 
Since this component was observed only during outburst phase, there must be a mechanism which is temporarily
available in outbursts. During quiescence, the mass accretion rate is not enough to create an optically thick 
boundary layer, but in outbursts there is observational evidence for such a boundary layer \citep{l1996}. 
Comparing the X-ray spectra obtained in quiescence and outburst we see that the optically thin plasma observed 
in quiescence, can also be seen in outburst with approximately the same temperature, but there exists an additional 
spectral component. Reflection of this X-ray emitting plasma from the optically thick boundary layer might occur 
in outburst giving rise to the extra X-ray component. Such reflection models are applied to AGNs, 
furthermore a similar detection for reflection during an outburst was reported by \shortcite{d1997} for SS Cygni. 
According to the reflection scenario, besides this extra component, one would expect to see $K_{\alpha}$ 
fluorescent lines \shortcite{b1995} and \shortcite{d1995}, due to the reflection of hard X-rays from the relatively 
cold EUV emitting boundary layer. Unfortunately in the high resolution Chandra spectra of U Gem during outburst 
those lines are either very weak or absent especially when compared to the resonance lines of Fe appearent near 
1.86 \AA. Moreover such reflection components are most effectively seen as a bump around 20 keV \citep{d1997}.  

Another plausible explanation for this hard X-ray component arises with the help of Low Inertia Magnetic Accretor 
model (LIMA) \citep{w2002}. 
According to the model, during outbursts the spin up of an accretion belt of the weakly magnetic white dwarf will 
enhance the intrinsic magnetic field and with this way, magnetically channelled accretion can occur even on low 
field CV primaries. With such an accretion onto the equatorial belt of the primary, accretion curtains and shocks in 
the same manner as in the standard intermediate polar structure are possible \citep{w2002}. This model is useful for 
explaining the dwarf nova oscillations (DNOs) observed in outbursts. In the model these DNOs are interpreted as the 
pulsations at the rotation period of this transient magnetosphere. 

As we have noted above, an equatorial accretion belt has been deduced from UV observations \citep{c1997}, \citep{l1993}. 
So if accretion curtains and shocks are likely to occur in U Gem during outbursts then a thermal bremsstrahlung or 
a power-law like emission may be observed from the cooling post-shock gas showing us an extra X-ray component. 
From the Chandra observations in quiescence (S02) and HST observations during outburst \citep{s1997} it has been 
suggested that magnetic accretion is possible for U Gem. 

We find a significant structure in one power spectrum that is similar to quasi periodic
oscillations (QPO). This feature appears around 73 mHz ($\sim 13.7$ s.).
Transient DNOs have been observed for U Gem, during outbursts, in the 25 s. range, by EXOSAT \citep{m1988} 
and EUVE \citep{l1996}. These oscillations may be arising from the transient magnetosphere which is predicted by the 
LIMA model. We, therefore, suggest that the extra X-ray spectral component as well as transient QPO we
detected originates from a transient magnetosphere. 

Another issue is the appearent changes in the emission line profiles, whose some of the properties are
summarized in Table \ref{emis_line}. Increase of the fluxes can be understood by assuming an high mass
transfer to the inner disc during outuburst. This will increase the probability of collisions 
between particles, which will of course increase collisional ionization.
However, explanation of the broadening of the lines can not be made as easy as increased fluxes, one assumption 
we can make is that the emission lines are now coming from a region rotating with Keplerian velocities, 
which influences the idea that the matter, at the outer parts of the accretion disc, moving towards to 
the white dwarf during the outburst. 
Although this idea is mainly in agreement with the standard outburst theories \citep{w1995}, it is not obvious from 
only one observation, because there are still some lines which are not as broadened as 
expected to form in a region rotating with a Keplerian velocity. 

In order to understand this phenomenon further broadband observations of U Gem and as well as other CVs, are needed to investigate the overall variations, particularly in the boundary layer, between the quiescence and outburst phases.

\section*{Acknowledgments}

Authors are grateful to the anonymous refree for useful suggestions. E. G. acknowledges partial support by the Turkish Academy of Sciences through grant E.G/TUBA-GEBIP/2004-11.
This work is made use of observations obtained with XMM-Newton, an ESA science mission with instruments and
contributions directly funded by ESA Member States and the USA (NASA). We acknowledge with thanks the variable
star observations from the AAVSO International Database, contributed by observers worldwide and used in this
research.

\label{lastpage}
\end{document}